\documentclass[10pt,twocolumn,letterpaper]{article}

\usepackage{cvpr}
\usepackage{times}
\usepackage{epsfig}
\usepackage{graphicx}
\usepackage{amsmath}
\usepackage{amssymb}

\usepackage[breaklinks=true,bookmarks=false]{hyperref}

\cvprfinalcopy 

\def\cvprPaperID{****} 

\ifcvprfinal\fi
\begin{document}
\title{\ 3DTI-Net:\ Learn Inner Transform Invariant 3D Geometry Features using Dynamic GCN}
\author{Guanghua Pan\\
SJTU\\
Shanghai\\
{\tt\small guanghua@sjtu.edu.cn}
\and
Jun Wang\\
SJTU\\
Shanghai\\
\and
Rendong Ying\\
SJTU\\
Shanghai\\
\and
Peilin Liu\\
SJTU\\
Shanghai\\
}
\maketitle

\begin{abstract}

Deep learning on point clouds has made a lot of progress recently. Many point cloud dedicated deep learning frameworks, such as PointNet and PointNet++, have shown advantages in accuracy and speed comparing to those using traditional 3D convolution algorithms. However, nearly all of these methods face a challenge, since the coordinates of the point cloud are decided by the coordinate system, they cannot handle the problem of 3D transform invariance properly. In this paper, we propose a general framework for point cloud learning. We achieve transform invariance by learning inner 3D geometry feature based on local graph representation, and propose a feature extraction network based on graph convolution network. Through experiments on classification and segmentation tasks, our method achieves state-of-the-art performance in rotated 3D object classification, and achieve competitive performance with the state-of-the-art in classification and segmentation tasks with fixed coordinate value.
\end{abstract}


\section{Introduction}

3D deep learning has achieved rapid progress recently, there are many different methods for 3D deep learning, some convey 3D shapes to voxel\cite{wu20153d}\cite{maturana2015voxnet:} while some project 3D objects to different image planes\cite{su2015multi-view}\cite{kanezaki2018rotationnet:}. However, the former represent the 3D data in a sparse way which is memory space consuming and ineffective, the latter relies on heavy computation and needs additional view rendering.

Point cloud learning has made a lot of progress and methods consuming point cloud directly as input have formed a new research field, which has shown advantages in accuracy and speed compared to other methods.  The first point cloud dedicated network is PointNet \cite{charles2017pointnet:}, which uses a transform net to align 3D shapes, and max pooling as global symmetric function to handle unordered point cloud. One weakness for PointNet is that it treats each point individually which leading to the loss of local structure information, Methods like PointNet++\cite{qi2017pointnet++:} and DGCNN \cite{wang2018dynamic} then try to solve this problem by taking neighbor points into account to improve the performance, but as the neighbor points in each patch are still treated individually, they fail to learn correlation of points within a patch.

Besides, since the coordinates of the points depends on the coordinate system, these methods require the orientation to be similar between training set and test set to achieve high accuracy in recognition tasks. 3D rotations have been a nuisance for 3D shape recognition tasks, most of point cloud based methods can only tackle it with data augmentation. Until recently some methods tries to solve the problem by projecting 3D points to a unit sphere and perform spherical CNN to learn $SO(3)$ equivalence 3D features \cite{cohen2018spherical}\cite{yi2017syncspeccnn:}. However these methods need preprocessing and are time consuming which limits the versatility of these methods.

We propose a transform-invariant 3D deep net taking a point cloud as direct input which we called 3DTI-Net, we designed a transform-invariant feature encoder and a deep net based on GCN(graph convolutional network) to learn transform-invariant 3D descriptor, Translation invariant and be easily achieved by re-centering the point cloud to the it's centroid, however, rotation invariant remains a challenge in 3D deep learning, Rotation invariant features are important in tasks like classification, segmentation or high-level task like 3D retrieval in practical application, for the reason that the point cloud collected by depth sensors may not aligned with the training data.

3DTI-Net can tackle rotation invariant problems without data augmentation and is more computationally efficient. Our model needs no pre-training and could consume point clouds directly without complex preprocessing.

The key contributions of our work are as follows:
(1) As far as we know, we are the first one to propose a graph-based transform-invariant 3D geometry feature for 3D point cloud learning, together with theoretical analysis on the transform-invariant property and the spatial meaning.

(2) We propose a novel, general GCN-based deep net to learn 3D geometry feature represented by graph signals more effectively.

(3) We propose a fast and effective transform-invariant feature based pooling method and a dynamic graph update strategy to achieve multi-resolution learning on graphs.

The rest of this paper is organized as follows: Related work is shown in section \uppercase\expandafter{\romannumeral2}. Problem statement of our work is given in section \uppercase\expandafter{\romannumeral3}. Details of our propose method is shown \uppercase\expandafter{\romannumeral4}. Experiment results are detailed in section \uppercase\expandafter{\romannumeral5}. Finally, we conclude this paper in section \uppercase\expandafter{\romannumeral6}.

\section{Related Work}

\textbf{3D Geometry Feature:}
3D feature descriptors could be split into two classes, hand-crafted features and learned features. Various  hand-crafted 3D features descriptors haver been proposed for segmentation, matching and classification, including spin images\cite{johnson1999using}, shape context\cite{belongie2000shape} and histogram based approaches\cite{rusu2009fast}\cite{tombari2010unique}\cite{rusu2008aligning}, most of which could be found in the Point Cloud Library\cite{aldoma2012tutorial:}. However, these methods have reached a bottleneck in handling noisy and low-resolution data, and those methods were designed for specific application which means they are hard to generalize for different situations.

Data-driven methods provide a better way with the success of CNN in 2D vision tasks. learned 3D features has achieved rapid progress since the introduction of 3D deep learning.

Su H \emph{et al.} \cite{qi2016volumetric} learns 3D shapes by a collection of the rendered views on 2D images and achieves dominating performance in 3D recongnition tasks. Zhou Y \emph{et al.} \cite{maturana2015voxnet:} propose VoxelNet which vonverts a point cloud into 3D voxels and transforms a group of points into a unified feature representation. However, both multi-view rending and voxelization lead to high time and space cost, besides, voxelization may cause an unnecessary information loss. Both methods above tried transforms point cloud to more regular data since point clouds are not in a regular format.

\textbf{Deep Learning on Point clouds:}
Charles R Q \emph{et al.} \cite{charles2017pointnet:} firstly design a novel deep network, named PointNet, for unordered point clouds. PointNet provides a unified architecture for various 3D based applications, such as object classification, part segmentation and scene semantic parsing. In order to better integrate local structure information, the authors propose another network called PointNet++ \cite{qi2017pointnet++:}. PointNet++ applies PointNet recursively on a nested partitioning of the input point set. Inspired by PoineNet, more researchers have propose different deep learning methods on point clouds \cite{wang2018dynamic}.

\textbf{Graph Signal processing:}
Graph signal processing are emerging field mean to combine signal processing and spectral graph theory, which aims at generalize fundamental signal operations from regular to irregular structures represented by graphs. we need new mathematical definitions to extend standard operations such as convolution, filtering, dilation and down sampling to graphs.

More research on graph signal processing tries to process data on irregular graph domains, readers could refer to \cite{Ortega2017Graph} for an overview, which summarizes recent development of basic graph signal processing tools, including methods for sampling, filtering and graph learning. Chen S \emph{et al.} \cite{chen2017fast} propose a re-sampling framework for 3D point clouds and a feature-extraction operator based on graph filtering. 

\textbf{Graph convolution neural network:}
\cite{bruna2014spectral} defined convolution over graphs in the spectral domain, however the computation complexity
is high due to the eigen-decomposition of the graph Laplacian matrix in order to get the eigenvector matrix. Defferrard M \emph{et al.} \cite{defferrard2016convolutional} then tries to impove GCN through fast localized convolutions, where Chebyshev expansion is deployed to approximate graph Fourier transform. 
 
Graph convolution networks have achieved great performance on learning data lying on irregular non-Euclidean domains like point clouds and social networks\cite{te2018rgcnn:}.

\textbf{Rotation Invariant 3D features Learning:}
Convolutional neural networks have translational invariance. However, 3D rotation equivariance in classical convolutional neural networks still remains unsolved. \cite{esteves2017learning}\cite{cohen2018spherical} propose a novel spherical convolutional network that implements exact convolution on the sphere. In order to achieve rotation equivariance, authors of \cite{thomas2018tensor} introduced a tensor field neural network that inputs and outputs tensor fields: scalars, vectors and higher-order tensors. \cite{worrall2018cubenet:} provided an example of a CNN with linear equivariance to 3D rotations and 3D translations of voxelized data.

\begin{figure*}[htp]
\centering
{
\includegraphics[width=1\linewidth]
                {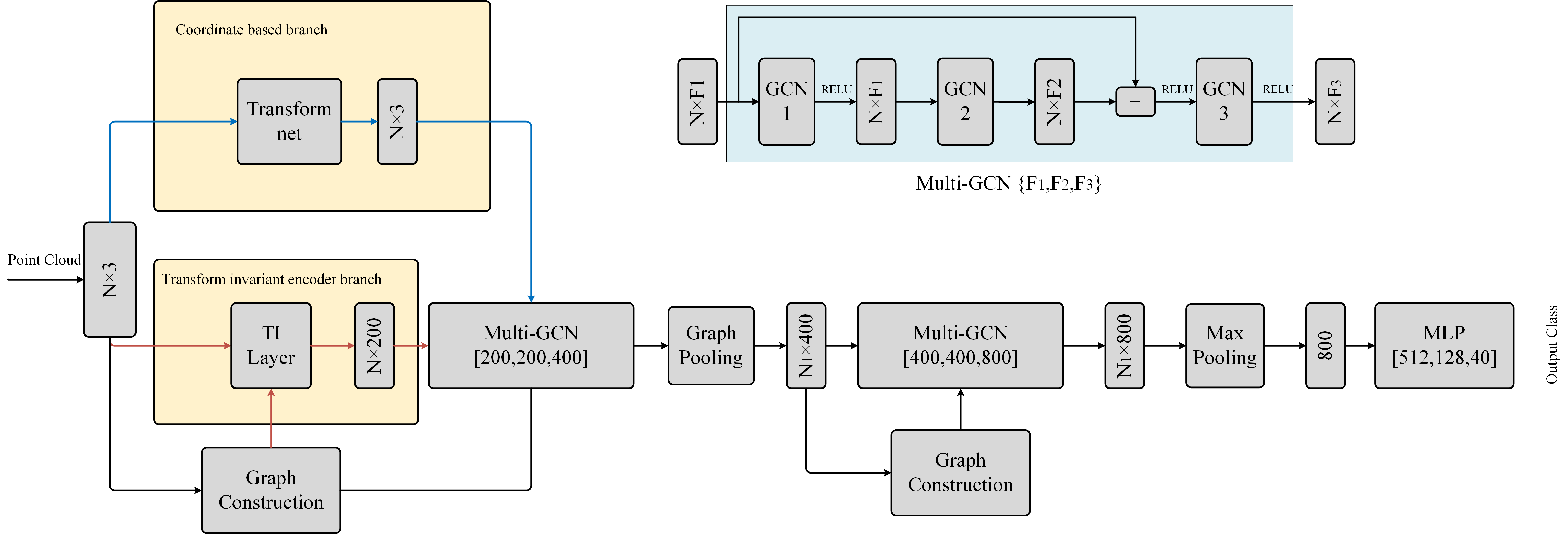}
\caption{Our propose model architecture, the left part is the front-end, which determines the input signal of the back end. Three modes
are considered: (1) trained and tested with azimuthal rotations (z/z), (2) trained and tested with arbitrary rotations (SO(3)/SO(3)), and (3) trained with azimuthal and tested with arbitrary rotations (z/SO(3)). If the test set and the training set are similar in distribution (z/z mode), we take the coordinates as direct input. if only test data are rotated (z/SO(3) mode), the original coordinates are transformed to high-dimensional transform invariant expression encoded by the TI-encoder. The right part is the back-end, which is composed of Multi-GCN layers, plays a role in extracting global feature.}
\label{model2}
}
\end{figure*}

\section{Problem Statement}

In order to propose a general network for point cloud learning, two challenges exits, the first one is to learn inner 3D geometry features in a transform-invariant way, the second one is to learn point cloud more generally and effectively with a deep net.
When given a 3D object $O$ represented by point cloud $X\in \Bbb{R}^{N\times C} $, where $N$ is the number of points and $C$ is number of feature channels, we need to design a transform-invariant feature encoder (abbreviated as TI-encoder) as the front-end to describe the shape of the 3D object, the encoded features should be able to express the inner geometric features. Let $H(.)\in {R}^{F_0}$ be the TI-encoder, where $F_0$ represents the dimension of the feature space after encoding, the encoded feature can be written as $X^{'}=H(X)$, Let $R$ be a random rotation matrix, then $X^{'}=H(RX)=H(X)$ should be satisfied to achieve a transform-invariant expression.
When given a point cloud $X\in \Bbb{R}^{N\times C}$ with $N$ points and each point with $C$ attributes, in classification experiments we use 3D coordinates or the encoded transform-invariant features to show that we can learn the geometry of a point cloud effectively, there are many attributes like colors, normal vectors or textures to further describe the characteristics of a point cloud, normals are used as additional attributes in segmentation experiments to better compare with previous methods. We need to design a feature extractor $F(.)\in \Bbb{R}^{N\times K}$ as the back-end to learn a feature descriptor, where $K$ is the length of the output feature descriptor. $F(.)$ should be able to extract features of high degree of division, high computational efficiency and be robust to noise and point cloud intensity.

\section{Method}

\subsection{Porposed Archetecture}

Our proposed model is shown in Figure \ref{model2}, the raw input of the model is point cloud $X\in \Bbb{R}^{N\times 3}$ with $N$ points and each with three coordinates, In section 4.2, the point cloud coordinates will be encoded as high dimensional transform-invariant feature $X'\in \Bbb{R}^{N\times F_0}$ through transform-invariant feature encoder(TI-layer), Multi-GCN layers are used to extract more abstract features, and in order to reduce the amount of parameters and better integrate local features, we perform pooling after the first Multi-GCN layer to achieve multi-resolution graph learning, the number of nodes is hence reduced and the information carried by each node becomes more abundant. Max pooling is applied to extract global feature descriptors, followed with three fully connected layers, which outputs the final predicted labels.

\subsection{3D Geometry Representation}

The coordinates of a point cloud depends on the choice of the 3D coordinate system. Therefore,
to achieve rotation invariant geometry feature learning, we need to convert the point cloud coordinates $X$ to another transform invariant expression $X'$, $X'$ should be able to express inner 3D geometry characteristic carried by the original 3D shape represented by the point cloud.

The geometric features are implicitly expressed by the relationship between different points. When the coordinate system where the point cloud located is fixed, a simple relationship between points can be represented as the difference in coordinate values $(\nabla{x},\nabla{y},\nabla{z})=(x_i-x_j,y_i-y_j,z_i-z_j) i,j\in N$, where $x_i,x_j$ are two points in point cloud $X$, representing the relative position in the coordinate system, from which we can get the distance information between two points: $l=norm(\nabla{x},\nabla{y},\nabla{z})$. Relative direction information in the determined coordinates $\Bbb{n}=(\nabla{x},\nabla{y},\nabla{z})/l$, and $(\nabla{x},\nabla{y},\nabla{z})$ can be completely determined by multiplying the length by a direction vector.

However, if the coordinate system is not fixed, the relative direction will change with the coordinate system while distance between points remain unchanged, functions based on distance are called RBFs(radial basis functions) which are widely used in handcrafted 3D rotation invariant features, one important property for RBFs is that they are rotation invariant. At the same time, if the distance between two points is known, the relationship between two points can be reconstructed when given a fixed coordinate system, and the geometric information carried by the two points can then be completely expressed. For a point cloud, if the pairwise distance between points are known, the inner geometric features carried by points can be completely reconstructed except for an additional mirror symmetry uncertainty.

Therefore, the geometric features can be implicitly represented by the connection relationship between points, the edge connecting each points are weighted by pair-wise distance. this expression depends only on the distance between points, hence is transform-invariant. The graph structure can be used to describe the relationship between points. All the points in point cloud serve as nodes. The weighted point-to-point connection relationship is the edge. There are many ways to express a graph. such as adjacency matrix $A$, Laplacian matrix $L$ and other descriptions.

\textbf{Graph signals}: We use the coordinates of the point as the origin signal, i.e. the signal of node $X_i$ is $x_i=[x_i^x,x_i^y,x_i^z]$, and $X\in \Bbb{R}^{N\times 3}$ means the whole signal consisting of all the nodes on the graph.

\textbf{Graph construction}: Graph can be regarded as a discrete approximation of the original surface. we use the distance in the original 3D space or in the feature space to construct a graph, it is also feasible to take additional feature into account. We could then define an undirected graphs $\mathcal{G}=(\nu, \epsilon, W)$, where $\nu $ is a finite set of $|\nu|=\Bbb{N}$ verticles, $\epsilon$ is a set of edges and $W \in \Bbb{R}^{n\times n}$ is a weighted adjacency matrix encoding the connection weight between verticles. First we calculate pairwise distances of the  point cloud, then find its $k$ neighbors according to the Euclidean distance of each point in 3D coordinate space to get the $k$ nearest neighbor adjacency distance matrix $ E\in \Bbb{R}^{N\times N}$, $E_{i,j}=\left\| x_i - x_j \right\|_2^2$ represents the Euclidean distance between the $i$th and $j$th point in the $N$ point, $N_i$ is the neighbor points of $i$th point, the weight between the center point and the neighbor point can then be computed based on the distance matrix:

\begin{equation}
\begin{split}
&W_{i,j} =
\left\{
    \begin{array}{rl}
      e^{- \frac{E_{i,j}}{\sigma^2} }, j\in Ni;\\
      0,  {otherwise},
  \end{array} \right.  \\
&\sigma={\frac{1} N}\times \sum_{i}^N \max\{{E_{i,j}}, j\in N_i.\}
\end{split}
\end{equation}

where $\sigma$ is the self-adaptive normalization parameter, and $W_{i,j}$ indicates the connection strength between two nodes, we can construct the adjacency matrix $A$ according to the weight. The symmetric normalized Laplacian matrix $L^{sys}=I-D^{-1/2}AD^{-1/2}$ can then be calculated where $D$ is the degree matrix of the adjacency matrix $A$. We also use random-walk normalized Laplacian matrix $L^{rw}=I-D^{-1}A$ as the input of our transform-invariant feature encode layer to achieve rotation invariance.

\subsection{Rotation Invariant Geometry Feature on Graphs}

We have constructed the graph structure in previous section, which is represented by the Laplacian matrix $L$, we also defined the signal $X$ on the graph.
Graph filtering is a method that takes a graph signal as raw input and outputs a processed graph signal. Let $S\in \Bbb{R}^{N\times N}$ be a graph shift operator, there are many common choices for $S$ in graph signal processing theory, such as adjacency matrix $A$ and Laplacian matrix $L=I-D^{-1}A$. When the graph shift operator acts on a node on the graph, the graph signal value of the central node is replaced by a weighted average of its neighbors, Each linear translation invariant graph filter kernel can be written as a polynomial of the graph shift operator \cite{sandryhaila2013discrete}:

\begin{equation}
\begin{split}
    h(S) \ &= \ \sum_{\ell = 0} ^{K-1} h_{\ell} S^{\ell} = h_0I + h_1S + \ldots + h_{K-1}S^{ K-1}
\end{split}
\end{equation}

Where $h_{\ell} (\ell = 0, 1, \ldots, K-1)$ represents the filter coefficient, K is the length of filter, and the output of the filtered signal $X$ is $y= h(S)X\in R^N$. let $S=L=I-D^{-1}A$, the filter kernel $L$ acts on the graph signal $X$, and the filtered output is $X^{'}=h(L)X=h_0IX + h_1LX + \ldots + h_{K-1}L^{K-1}X$.

\textbf{Theorem1}: Let $X$ be the coordinates of a point cloud recentered to it's centroid, define $K$ features on graph as follows:
$f_i (X) = \left\| \ \left( h(L) X \right)_i \right\|_2^2,i\in 0,\dots,K-1$, 
Where $\left\| \ \left( h(L) X \right)_i \right\|_2^2$ represents $(h(L) X)_i$ norm of each row, then the $f_i (X)$ is rotational and translation invariant.

\textbf{Proof}:
Translation invariant can be easily achieved by the re-centering operation, To prove rotation invariance, let ${R} \in \Bbb{R}^{3 \times 3}$ be a random rotation matrix, and the rotated point cloud coordinates is $XR$. Then the $i$ $_th$ feature:

%
\begin{equation}
\begin{split}
f_i (X R ) & =  \left\|  \left( h(L) X R \right)_i \right\|_2^2 \\
        & =    \left\|  \left( h(L)  \right)_i  X R \right\|_2^2 \\
        & =   \left( h(L)  \right)_i  X R  R^T X^T \left( h(L)  \right)_i ^T \\
        & =   \left( h(L)  \right)_i  X X^T \left( h(L)  \right)_i ^T \\
        & =   \left\|  \left( h(L) X \right)_i \right\|_2^2 \ = \ f_i ( X )
\end{split}
\end{equation}

%
\textbf{TI (Transform-Invariant) Layer:}

Given a graph represented by Random Walk Laplacian matrix $L^{rw}=I-D^{-1}A$ together with graph signal $X\in \Bbb{R}^{N\times 3}$, the filtered graph signal is $X^{'}\in \Bbb{R}^{N\times F_0}$, where $F_0$ is the number of feature map, the polynomial order is $K$, then the $j$ th output channel of TI layer can be write as:

${X_j}^{'}=\sum_{i=0}^{K-1} \theta_{ij} \left\|(L^{rw})^{i}X)\right\|_2^2+b_j, j\in 0,1,...,F_0-1$

where $b_j$ are the bias of $j$ $_th$ channel of the feature map, $\theta_{0j},...,\theta_{(K-1)j}$ are K trainable filter coefficients which can be regarded as the convolution kernel.

\textbf{Transform invariant feature visualization:}

\begin{figure}
{
\includegraphics[width=1\linewidth]
                {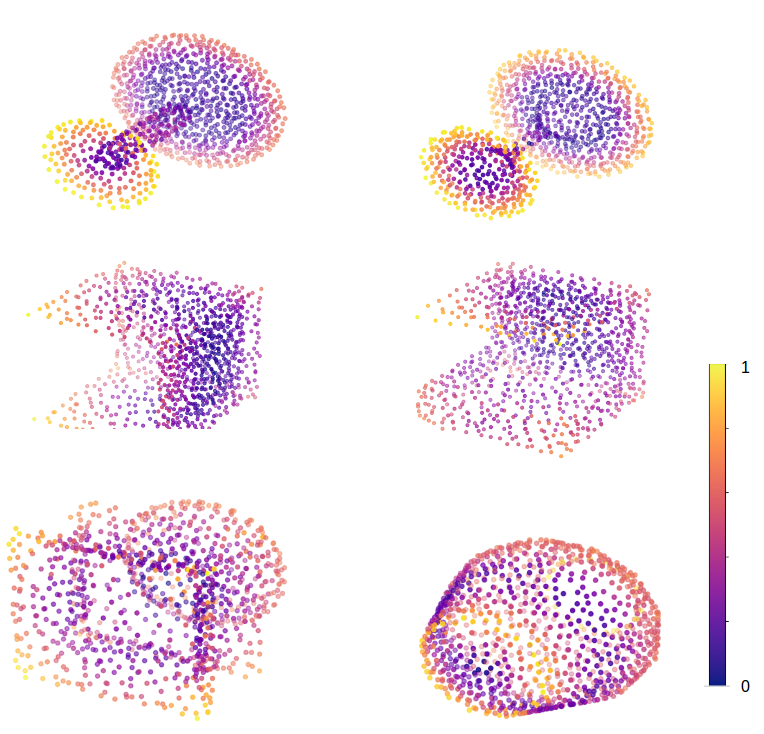}
\caption{Each row in the image corresponds to a category, and each column of a row corresponds to two samples in the same class, We have normalized the $l_2$ length of the transformed features to 1.}
\label{ti-feature}
}
\end{figure}

The visualization result of point cloud features after transform invariant encoding. is shown in Figure \ref{ti-feature}, It can be seen that the value is associated with contours of 3D object.

\subsection{Spatial interpretation of Rotation Invariant feature}

As mentioned in section 4.2, when applying the graph shift operator on a point, the operation on the spatial domain is calculating the weighted average of the neighbor points around the center point. For Random Walk Laplacian matrix $L^{rw}=I-D^{-1}A$, the sum of each row equals to $0$, so $L^{rw}X =X-D^{-1}AX$ can be be decomposed into two parts: $D^{-1}AX$ calculates the weighted average of the neighbor points around the central point. the weight is calculated by the distance between neighbor points and the center point. $X-D^{-1}AX$ calculate the the relative coordinates of center point to the weighted average point.
Therefore $\left\|L^{rw}X\right\|$ represents the Euclidean distance from the center point to the weighted average point, it's a more comprehensive understanding of its rotation invariance property since the Euclidean distance is an invariant variable in $SO (3)$. We call this distance contour variance.
The rotation invariantce can be achieved by calculating the norm of the relative coordinates, but the direction information is ignored. let $T=L^{rw}X$ be a relative coordinate, and the coordinate value of any point is $T_i=(x'_i, y'_i, z'_i)$. the direction vector of a each point can be defined as $n_i=T_i/\left\|(T_i)\right\|_2^2)$ which is also associated with the choice of coordinate system. The direction of a single node is also meaningless, the geometric information still hidden in the orientation relationship between points.
Thus we can treat the direction vector as a new graph signal and perform feature extraction operation in the same way, that is, we apply a graph filtering to the new graph signal to obtain a description of the direction variance between each points. The feature is also rotation invariant and can be easily proved in the same way, we call this new feature direction variance, we then concatenate contour variance and direction variance to get the final output of TI layer.

\subsection{Rotation Invariant convolution on graphs}

In section 4.2, the point coordinates $X\in \Bbb{R}^{N\times 3}$, which will change with rotation, are converted into a high-dimensional rotation-invariant feature $X^{'}\in \Bbb{R}^{N\times F_0}$ by the front-end. that is, The absolute 3D coordinate value can then be transformed to relative rotation invariant feature through the relationship between points. we still need to design a network to extract deep rotation invariant feature.

GCN is used to extract feature descriptor of the point cloud. the convolution operator on a graph is originally defined in the spectral domain \cite{bruna2014spectral}, however, two limitations exits:
(1) it is not localized in space, (2) due to the eigen-decomposition of the graph Laplacian, it has high computational complexity. Defferrard \emph{et al.} propose to use Chebyshev polynomials to approximate the spectral filtering \cite{defferrard2016convolutional}. The K-localized filtering operation can be defined as:

\begin{equation}
\begin{split}
\mathbf{y} = g_\theta(\mathcal{L}) \mathbf{x}
            = \sum_{k=0}^{K-1}\theta_kT_k(\mathcal{L}) \mathbf{x}
\end{split}
\end{equation}
$\theta_k$ is the $k$-th Chebyshev coefficient. $T_k(\mathcal{L})$ is the Chebyshev polynomial of order $k$. It can recurrently calculated by $T_k(\mathcal{L}) = 2\mathcal{L}T_{k-1}(\mathcal{L}) - T_{k-2}(\mathcal{L})$, where $T_0(\mathcal{L}) = 1,T_1(\mathcal{L}) = \mathcal{L}$. Then the complexity is reduced from $\mathcal{O}(N^3)$ to $\mathcal{O}(K |\mathcal{E}|)$.

K-localized GCN is used as our convolution layer. The output of the TI layer is transform invariant: $X^{'}=f(X)=f(RX+t), X^{'}\in R^ {N\times F_0}$ where $R,t$ are arbitrary rotation and translation matrices. So the entire convolution process is rotation invariant:

\begin{equation}
\begin{split}
{y} &= g_\theta(\mathcal{L}) {f(X)} = g_\theta(\mathcal{L}) {f(RX+t)}\\
    &=\sum_{k=0}^{K-1}\theta_kT_k(\mathcal{L}) {X^{'}}\\
\end{split}
\end{equation}

We then use a relu activation function to achieve the nonlinearity of the network:

\begin{equation}
y = \text{ReLU}( g_\theta(\mathcal{L})X^{'}\mathbf{W}+ {b}),
\end{equation}
where $W\in \Bbb{R}^{F_0\times F_1}$ are trainable weights, $b\in \Bbb{R}^{N\times F_1}$ are the bias, $y\in \Bbb{R}^{N\times F_1}$ is the final output.

\subsection{Multi-resolution learning on graphs}

\begin{figure}[h]
{
\includegraphics[width=1\linewidth]
                {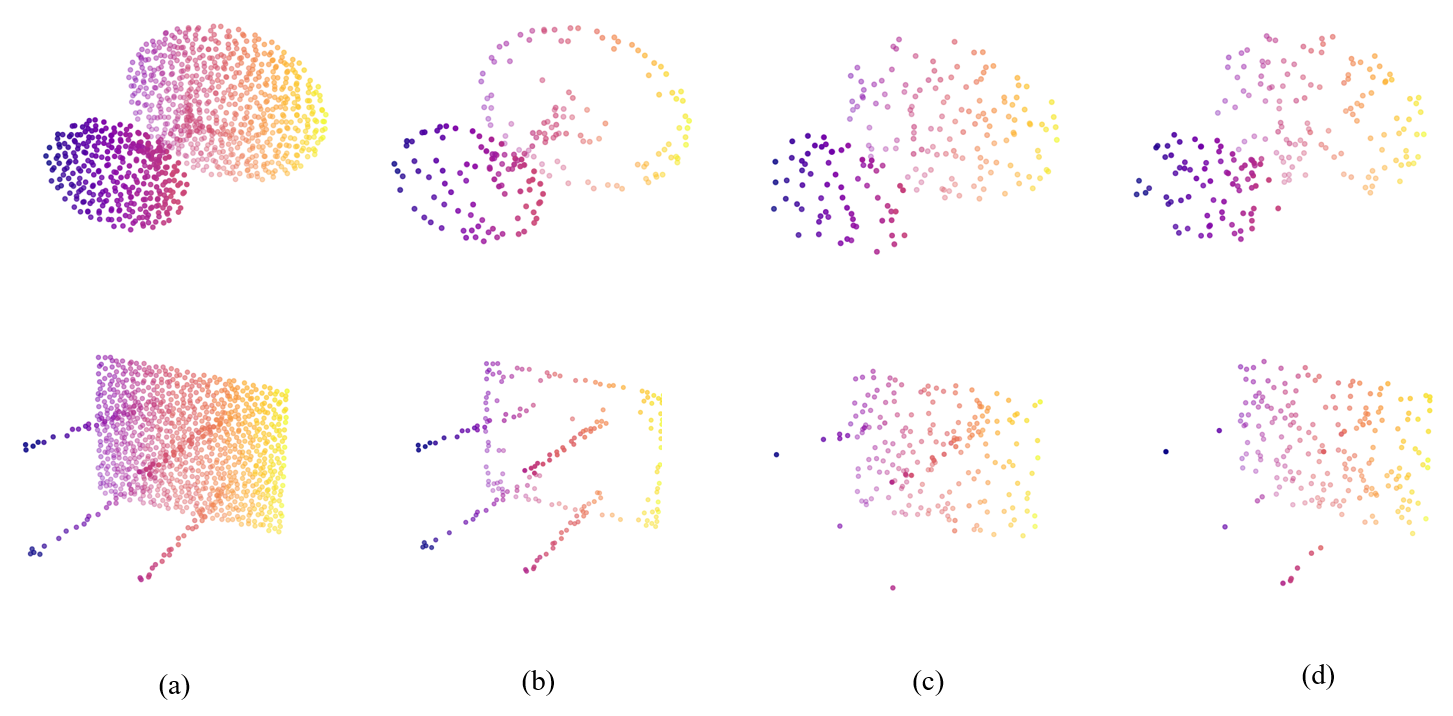}
\caption{Coarsening on graphs, column (a) is the origin point cloud, (b) is the proposed down sampling method, (c) is uniform down sampling while (d) is furthest point sampling}
\label{pooling}
}
\end{figure}

\textbf{Fast transform-invariant pooling on graphs:}
The pooling operation consists of two steps. first we need to down sample the point cloud and keep those points that contributes more on final classification result after coarsening, they are similar to the 'critical points' in PointNet which are usually the contour points of an object. Then we need to cluster points that are most similar to the sampled points.
Coarsening on point cloud is a challenging task, some methods simply treat each point equally and use uniform down sampling to handle this problem, but uniform sampling suffers uncertainty and the sampled points does not contributes more on the final result, others \cite{10} use furthest points sampling but still have the same problem and is far more time consuming. pooling is frequently used in deep net so pooling on point cloud must be computation effective too.
The TI feature we propose is related to the contour variance of the point, i.e. the difference between the center point and it's neighbor points and we have explained the spatial meaning of it in section, Just like the edges in 2D images, we think the contours of a point clouds are points that are different from their neighbor points, so we just sort the TI feature and select top $N'$ points that vary most from their neighbors. the comparison of our coarsening strategy with uniform sampling is show in \ref{pooling}
After coarsening, we search $m$ nearlist neighbors of each sampled point, followed with a local maxpooling to integrate local information, $m$ is set to 8 in our experiment.

\textbf{Dynamic Graph Updating:}
The dynamic graph has three meanings, first, the number of neighbors $k$ used to construct the graph is dynamic in order to adjust the receptive fields of different layers. Second, in classification task, the number of nodes of the graph is dynamic since the amount of nodes is reduced after pooling. Third, the connection relationship and the weight of the edge are determined by the distance in the feature space in segmentation task.

\subsection{3DTI-Net Baseline Model}

\begin{figure}
{
\includegraphics[width=1\linewidth]
                {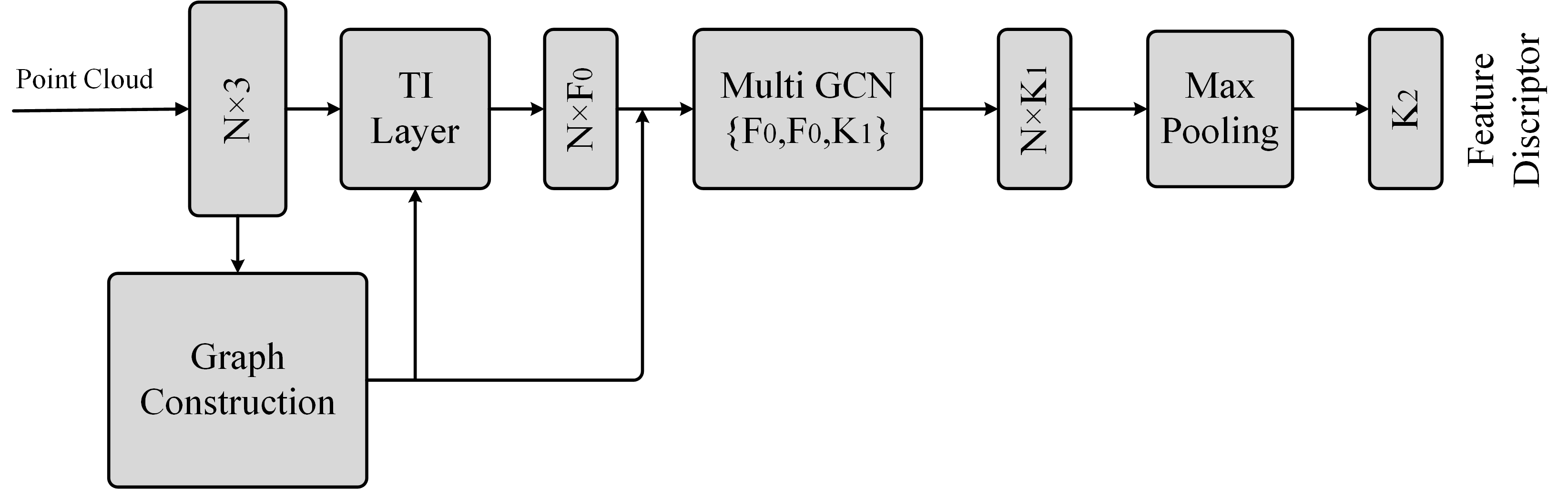}
\caption{Our baseline model, the coordinates are transfered to a high-dimentional transform invariant representation through TI-layer, then two GCN layers and a maxpooling are used to output final feature descriptor.}
\label{model1} 
}
\end{figure}

To show that 3DTI-Net is effective in learning inner 3D geometry features, a simple network architecture is designed as our baseline model. The baseline mainly consists of 3 parts, the translation invariant geometry feature encoder layer, feature extraction layer based on graph CNN, and a global feature extraction layer.
The point cloud coordinates $X$ can be encoded as high dimensional transform-invariant feature $X'\in \Bbb{R}^{N\times F_0}$ through rotation invariant feature encoder layer, then pass $X'$ through feature extract layer which is composed of 2 GCN layers to learn feature that represents structure information, a golbal max pooling is then applied to extract global rotation invariant 3D geometry feature descriptor.

\section{Experiment}

\subsection{Transform Invariance Test}

The transform invariance property of our network are tested through 3D shape classification task and 3D retrieval task. For classification task, when given a set of 3D objects with label, we need to train a function $l=f(X)$ to predict the label of a new object. We choose ModelNet40 as our dataset which is widely used in 3D object classification tasks. ModelNet40 contains 12311 CAD models from 40 categories, for each model, 1,024 points are uniformly sampled from the mesh faces and normalized to the unit sphere. We only use the coordinates of the sampled points in classification task.
We train the model with the origin oriention and test the data with random rotation to see whether our model could achieve translation invariant in classification task (z/SO(3) mode).

\textbf{Implementation Details}

There exits two strategies to avoid over-fitting, we add dropout after fully connected layer and add regularization term to total loss, we also use a weighted gradient descent to prevent the model from over fitting to classes with majority samples. The models are trained using an NVIDIA TITAN GPU, with a batch size of 16 and epoch number of 400. 

\begin{table}
  \caption{ModelNet40 test result, overall classification accuracy. \label{tab:m40}}
  \centering
    {\def\arraystretch{1}\setlength\tabcolsep{4pt}
    \begin{tabular}{lcccl}
      \hline
      Method  & SO3/SO3 & z/SO3  & inp. size & Para\\
      \hline
      PointNet++ \cite{qi2017pointnet++:}  & 85.0 & 28.6 & $1024 \times 3$ &1.7M\\
      SubVolSup MO \cite{qi2016volumetric}  & 85.0 & 45.5 &  $20 \times 30^3$ &17M\\
      MVCNN 12x \cite{su2015multi-view}  & 77.6 & 70.1 &  $12 \times 224^2$ &99M\\
      MVCNN 80x \cite{su2015multi-view} & 86.0 & - & $80 \times 224^2$ &99M \\
      RotationNet 20x \cite{kanezaki2018rotationnet:}  & 80.0 & 20.2 & $20 \times 224^2$ &58.9M\\
      Sperical-CNN \cite{esteves2017learning} & {86.9} & {78.6} & ${2 \times 64^2}$ &0.5M\\
      \hline
      Ours baseline & -  & {81.6} & ${1024 \times 3}$ &\textbf{0.4M}\\
      Ours  & \textbf{87.1} & \textbf{83.9} & ${1024 \times 3}$ &4.8M\footnotemark[1]\\
      \hline
    \end{tabular}
    }
\end{table}
\footnotetext[1]{The model size of mode z/SO(3) is 2.6M}
We compared our methods with state-of-the-arts, our proposed method outperforms previous state-of-the-art Sperical-CNN by a gap 5.3\%. Previous coordinate based methods \cite{kanezaki2018rotationnet:,qi2017pointnet++:} could achieve good performance when the orientation of the training data and the test data are aligned, but there is a sharp drop in accuracy when the test data are randomly rotated. Multi-View based methods could be regard as a brute force approach of $SO(3)$ equivariance, they could generalizes to unseen orientations\cite{esteves2017learning}, but still, our method outperform them with a magnitude fewer parameters and no pre-training.
Previous point cloud based methods have tried to handle rotation invariance by data argumentation, which require higher model capacity. As shown in Section 5.2, the back-end of 3DTI-Net is also capable of learning the coordinates directly to achieve higher performance in recognition tasks. We trained an additional model with data argumentation i.e. using the rotated training set, and then we concatenate the descriptors learned by two models and trained a voting network to output the final result, we achieve state-of-the-art prediction accuracy when the training set and test set are all randomly rotated (SO(3)/SO(3) mode).

\subsubsection{3D Retrieval}

We perform retrieval experiment on ShapeNet Core55, which is composed of more than 50 thousands models
over 55 common categories in total for training and evaluating, In our experiment each query and retrieval results are treated equally across categories, and therefore the results are averaged without re-weighting based on category size.
We use the same model architecture shown in Figure \ref{cls}, we follow the setting of \cite{esteves2017learning}, the network are trained for classification on 55 core classes with an additional loss to minimize the distance between matching categories textbf{while pulling apart non-matching categories at the same time}, we use the distance between descriptors for retrieval. the elements whose distance are below the threshold are returned. 
As shown in \ref{retrieval}, our method achieve the best trade-off between the model complexity, input size and retrieval performance. The result also proves that 3DTI-Net are effective in learning inner features of 3D objects, and the learned transform invariant features can be widely used in matching and recognition tasks.

\begin{table}[h]
  \caption{3D retrieval result in SHREC'17, we compare precision,recall and mean average precision.\label{retrieval}
}%
  \centering
  {\def\arraystretch{1}\setlength\tabcolsep{4pt} 
  \begin{tabular}{l|rrrll}
    \hline
    mothod                                & P@N                        & R@N        & mAP &input size &params         \\
    \hline
    Furuya \cite{furuya2016deep}    & \textbf{0.814}                 & 0.683      & 0.656   & $126\times 10^3$         & 8.4M     \\
    Tatsuma \cite{tatsuma2009multi} & 0.705                      & \textbf{0.769} & 0.696 &$38\times224^2$          & 3M \\
    Zhou \cite{bai2016gift}         & 0.660                      & 0.650      & 0.567   &$50\times224^2$   & 36M    \\
Sperical-CNN \cite{esteves2017learning}               & 0.717                & 0.737      & 0.685  & ${2\times64^2}$   & 0.5M\\
    \hline
    ours                                                        & 0.749            & 0.756 & \textbf{0.698} & \textbf{$1024\times 3$} &2.6M\\
    \hline
  \end{tabular}
  }
\label{tab:shrec}
\end{table}

\subsection{Back-end Capability Test}
Our model is effective in learning 3D transform invariant feature but it's not limited to this condition, in this section we will show that our model is a general framework in learning 3D geometry feature represented by graph signal, we train and test the model without rotation just like most previous point cloud based models did, we consider two prototypical tasks in this field: classification and segmentation. We perform classification experiment on Modelnet40 and segmentation experiment on ShapeNet to test the capability of our model. 

\subsubsection{Classification}
\textbf{Architecture }
The network architecture is illustrated in Figure \ref{model2}. we follow previous point cloud based methods \cite{charles2017pointnet:,qi2017pointnet++:,wang2018dynamic} train and test without rotation in ModelNet40, the result is shown in Table \ref{cls}.

\begin{table}[h]
  \caption{Classification task, test with coordinates, we compare our method overall classification accuracy forward time and model size . \label{cls}}
  \centering
    {\def\arraystretch{1}\setlength\tabcolsep{5pt}
    \begin{tabular}{lcccc}
      \hline
      Method  & overall acc&fwd time(MS) & model size\\
      \hline
      PointNet \cite{charles2017pointnet:} & 89.2 &25.3 & 3.5M \\ 
      PointNet++ \cite{qi2017pointnet++:} & 90.7 & 163 & 1.7M\\
      RotationNet 20x \cite{kanezaki2018rotationnet:} & -& 92 & 58.9M\\
      DGCNN\cite{wang2018dynamic} & \textbf{92.2} & 94.6 & 2.1M\\
      \hline
      Ours baseline & 89.5 & \textbf{10.8} & \textbf{0.4M} \\
      Ours & 91.7 & 60 & 2.6M\\
      \hline
    \end{tabular}
    }
\end{table}
\begin{figure*}[h]
{
\includegraphics[width=1\linewidth]
                {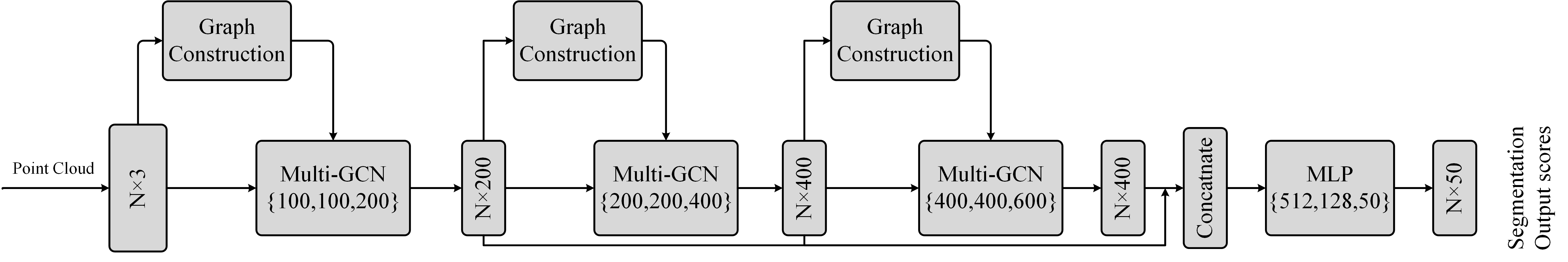}
\caption{model for segmentation.}
\label{seg_model} 
}
\end{figure*}
Though there is still a gap between our method and multi-view based methods, we achieve competitive performance with previous state-of-the-arts which comsume point cloud directly, while our outperform them by fewer parameters and more effective in computation.

\begin{figure}
{
\includegraphics[width=1\linewidth]
                {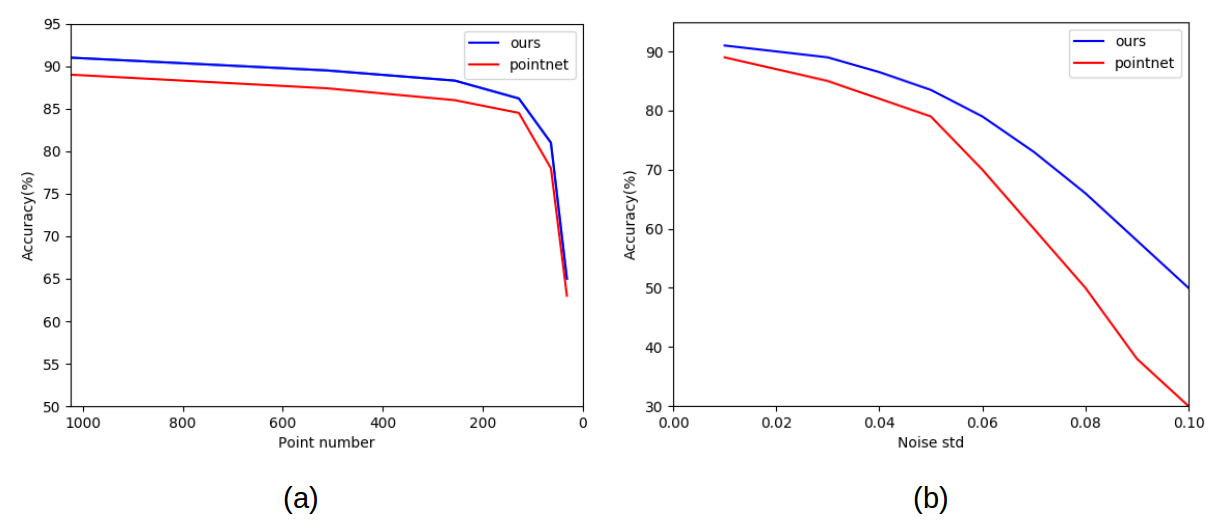}
\caption{Sensitivity to point intensity}
\label{pressure}
}
\end{figure}

\textbf{Pressure test:}
We jitter the coordinates of the point cloud with Gauss noise $n$, we set the mean of $n$ to zero, then we increase standard deviation $\sigma$ from 0.01 to 0.1 to see how the prediction accuracy changes, the result is shown in Figure \ref{pressure}, the result shows that our model is robust to noise.

We train and test using point cloud with different intensity, we test the prediction accuracy and compare our result with PointNet, as shown in Figure \ref{pressure}, our model is more stable then PointNet and keeps $88.3 \%$ accuracy even with only 256 points.

\subsubsection{Segmentation}
Point segmentation is classifying the label of each point in a 3D point set, which is more challenging than point set classification. We present experiment on ShapeNet, which contains 16881 shapes from 16 categories and annotated with 50 labels. In the experiments, we first  extract 2048 points from each model with random sampling, most sampled point sets have less than six labeled parts. We then feed the coordinates and normal as raw input to our model.

As shown in Figure \ref{seg_model}, the input is a point cloud with coordinates and normal as attributes, we then construct a graph represented with Laplacian matrix, we set the nearest neighbor $K$ to 200 to better capture local geometry feature, we pass the graph signal through a GCN layers, then we dynamicly update the graph using the distance in feature space and raise the nearest neighbor number search $K$ to 1000, in last Multi-GCN layer, we use fully connected graph to better integrate global information, finally we concatanate the output feature of three Multi-GCN layers and use a fully connected layer to predict the final output segmentation scores.

We assume that a category label for each shape is already known and use the mIoU (mean intersection over union) metric to eveluate the result. We compare our method with state-of-the-art approaches, as shown in Table \ref{seg}, our method outperform PointNet and achieves competitive performance with the state-of-the-arts.

We also performed segmentation experiment in $SO(3)$ with transfrom-invariant feature, we achieve decent performance while other methods suffer sharp drops in mIOU, that's mainly because previous methods relies heavily on the coordinates instead of the inner 3D geometry feature implicitly represented by the correlation between points.

\begin{table}
  \caption{Segmentation with coordinate, mIOU. \label{seg}}
  \centering
    {\def\arraystretch{1}\setlength\tabcolsep{5pt}
    \begin{tabular}{lc}
      \hline
      Method & mIOU \\
      \hline
      PointNet \cite{charles2017pointnet:} & 83.7 \\
      PointNet++ \cite{qi2017pointnet++:} & 85.1 \\
      SynSpecCNN \cite{yi2017syncspeccnn:} & 84.7 \\
      DGCNN\cite{wang2018dynamic} &85.1 \\
      \hline
      Ours & 84.9 \\
      \hline
    \end{tabular}
    }
\end{table}

\section{Conclusion}

In this work, we propose, a novel general deep neural network 3DTI-Net for point cloud learning, which could not only learn shape represented by coordinates but also learn inner geometry features represented by correlation between points. We propose a learned 3D transform invariant geometry feature based on graph, we show its performance on various tasks, the success of our model suggests that intrinsic features are also very important, for a point cloud, the intrinsic geometry lies on the correlation between points, which can be easily represented by a graph. And the geometry represented by graph can be learned using GCN, the features we learn are spatially localized since the graph filters with length $K$ combine information strictly from K-hop neighbors, which are encoded in the constructed graph. 3DTI-Net consume point cloud directly with no preprocessing, pre-training which makes the method suitable for various application scenarios. The model has a small amount of parameters and is computational efficiency. These characteristics make the method suitable for real-time applications, such as semantic slam.

{
\bibliographystyle{ieee}
\bibliography{reference}

\begin{thebibliography}{10}\itemsep=-1pt

\bibitem{aldoma2012tutorial:}
A.~Aldoma, Z.~Marton, F.~Tombari, W.~Wohlkinger, C.~Potthast, B.~Zeisl, R.~B.
  Rusu, S.~Gedikli, and M.~Vincze.
\newblock Tutorial: Point cloud library: Three-dimensional object recognition
  and 6 dof pose estimation.
\newblock {\em IEEE Robotics \& Automation Magazine}, 19(3):80--91, 2012.

\bibitem{bai2016gift}
S.~Bai, X.~Bai, Z.~Zhou, Z.~Zhang, and L.~J. Latecki.
\newblock Gift: A real-time and scalable 3d shape search engine.
\newblock In {\em Computer Vision and Pattern Recognition}, pages 5023--5032,
  2016.

\bibitem{belongie2000shape}
S.~J. Belongie, J.~Malik, and J.~Puzicha.
\newblock Shape context: A new descriptor for shape matching and object
  recognition.
\newblock pages 831--837, 2000.

\bibitem{bruna2014spectral}
J.~Bruna, W.~Zaremba, A.~Szlam, and Y.~Lecun.
\newblock Spectral networks and locally connected networks on graphs.
\newblock {\em international conference on learning representations}, 2014.

\bibitem{charles2017pointnet:}
R.~Q. Charles, H.~Su, M.~Kaichun, and L.~J. Guibas.
\newblock Pointnet: Deep learning on point sets for 3d classification and
  segmentation.
\newblock {\em computer vision and pattern recognition}, pages 77--85, 2017.

\bibitem{chen2017fast}
S.~Chen, D.~Tian, C.~Feng, A.~Vetro, and J.~Kovacevic.
\newblock Fast resampling of 3d point clouds via graphs.
\newblock {\em arXiv: Computer Vision and Pattern Recognition}, 2017.

\bibitem{cohen2018spherical}
T.~S. Cohen, M.~Geiger, J.~Kohler, and M.~Welling.
\newblock Spherical cnns.
\newblock {\em international conference on learning representations}, 2018.

\bibitem{defferrard2016convolutional}
M.~Defferrard, X.~Bresson, and P.~Vandergheynst.
\newblock Convolutional neural networks on graphs with fast localized spectral
  filtering.
\newblock {\em neural information processing systems}, pages 3844--3852, 2016.

\bibitem{esteves2017learning}
C.~Esteves, C.~Allenblanchette, A.~Makadia, and K.~Daniilidis.
\newblock Learning so(3) equivariant representations with spherical cnns.
\newblock {\em arXiv: Computer Vision and Pattern Recognition}, 2017.

\bibitem{furuya2016deep}
T.~Furuya and R.~Ohbuchi.
\newblock Deep aggregation of local 3d geometric features for 3d model
  retrieval.
\newblock In {\em British Machine Vision Conference}, pages 121.1--121.12,
  2016.

\bibitem{johnson1999using}
A.~E. Johnson and M.~Hebert.
\newblock Using spin images for efficient object recognition in cluttered 3d
  scenes.
\newblock {\em IEEE Transactions on Pattern Analysis and Machine Intelligence},
  21(5):433--449, 1999.

\bibitem{kanezaki2018rotationnet:}
A.~Kanezaki, Y.~Matsushita, and Y.~Nishida.
\newblock Rotationnet: Joint object categorization and pose estimation using
  multiviews from unsupervised viewpoints.
\newblock {\em computer vision and pattern recognition}, 2018.

\bibitem{maturana2015voxnet:}
D.~Maturana and S.~Scherer.
\newblock Voxnet: A 3d convolutional neural network for real-time object
  recognition.
\newblock pages 922--928, 2015.

\bibitem{Ortega2017Graph}
A.~Ortega, P.~Frossard, J.~Kovacevic, J.~M.~F. Moura, and P.~Vandergheynst.
\newblock Graph signal processing: Overview, challenges, and applications.
\newblock {\em Proceedings of the IEEE}, 106(5):808--828, 2017.

\bibitem{qi2016volumetric}
C.~R. Qi, H.~Su, M.~Niebner, A.~Dai, M.~Yan, and L.~J. Guibas.
\newblock Volumetric and multi-view cnns for object classification on 3d data.
\newblock {\em computer vision and pattern recognition}, pages 5648--5656,
  2016.

\bibitem{qi2017pointnet++:}
C.~R. Qi, L.~Yi, H.~Su, and L.~J. Guibas.
\newblock Pointnet++: Deep hierarchical feature learning on point sets in a
  metric space.
\newblock {\em neural information processing systems}, pages 5099--5108, 2017.

\bibitem{rusu2009fast}
R.~B. Rusu, N.~Blodow, and M.~Beetz.
\newblock Fast point feature histograms (fpfh) for 3d registration.
\newblock pages 1848--1853, 2009.

\bibitem{rusu2008aligning}
R.~B. Rusu, N.~Blodow, Z.~Marton, and M.~Beetz.
\newblock Aligning point cloud views using persistent feature histograms.
\newblock pages 3384--3391, 2008.

\bibitem{sandryhaila2013discrete}
A.~Sandryhaila and J.~M.~F. Moura.
\newblock Discrete signal processing on graphs.
\newblock {\em IEEE Transactions on Signal Processing}, 61(7):1644--1656, 2013.

\bibitem{su2015multi-view}
H.~Su, S.~Maji, E.~Kalogerakis, and E.~G. Learnedmiller.
\newblock Multi-view convolutional neural networks for 3d shape recognition.
\newblock {\em international conference on computer vision}, pages 945--953,
  2015.

\bibitem{tatsuma2009multi}
A.~Tatsuma and M.~Aono.
\newblock Multi-fourier spectra descriptor and augmentation with spectral
  clustering for 3d shape retrieval.
\newblock {\em Visual Computer}, 25(8):785--804, 2009.

\bibitem{te2018rgcnn:}
G.~Te, W.~Hu, A.~Zheng, and Z.~Guo.
\newblock Rgcnn: Regularized graph cnn for point cloud segmentation.
\newblock {\em acm multimedia}, 2018.

\bibitem{thomas2018tensor}
N.~Thomas, T.~Smidt, S.~Kearnes, L.~Yang, L.~Li, K.~Kohlhoff, and P.~F. Riley.
\newblock Tensor field networks: Rotation- and translation-equivariant neural
  networks for 3d point clouds.
\newblock {\em arXiv: Learning}, 2018.

\bibitem{tombari2010unique}
F.~Tombari, S.~Salti, and L.~D. Stefano.
\newblock Unique signatures of histograms for local surface description.
\newblock pages 356--369, 2010.

\bibitem{wang2018dynamic}
Y.~Wang, Y.~Sun, Z.~Liu, S.~E. Sarma, M.~M. Bronstein, and J.~Solomon.
\newblock Dynamic graph cnn for learning on point clouds.
\newblock {\em arXiv: Computer Vision and Pattern Recognition}, 2018.

\bibitem{worrall2018cubenet:}
D.~E. Worrall and G.~J. Brostow.
\newblock Cubenet: Equivariance to 3d rotation and translation.
\newblock {\em arXiv: Computer Vision and Pattern Recognition}, 2018.

\bibitem{wu20153d}
Z.~Wu, S.~Song, A.~Khosla, F.~Yu, L.~Zhang, X.~Tang, and J.~Xiao.
\newblock 3d shapenets: A deep representation for volumetric shapes.
\newblock {\em computer vision and pattern recognition}, pages 1912--1920,
  2015.

\bibitem{yi2017syncspeccnn:}
L.~Yi, H.~Su, X.~Guo, and L.~J. Guibas.
\newblock Syncspeccnn: Synchronized spectral cnn for 3d shape segmentation.
\newblock {\em computer vision and pattern recognition}, pages 6584--6592,
  2017.

\end{thebibliography}
}
\end{document}